\documentclass[journal,twoside]{IEEEtran}

\usepackage{float}
\usepackage{stackengine}
\usepackage{setspace}
\usepackage{amssymb,amsfonts,latexsym}
\usepackage{graphicx}
\usepackage{subfigure}
\usepackage{psfrag}
\usepackage{algorithmicx}
\usepackage{blindtext}
\usepackage{comment}
\usepackage{cite}
\usepackage{url}
\usepackage{times}
\usepackage{bbm}
\usepackage{epsfig}
\usepackage{multirow}
\usepackage{longtable}
\usepackage{amsfonts}
\usepackage{amssymb}%
\usepackage{color}
\usepackage{mathrsfs}
\usepackage{bm}
\usepackage{booktabs}
\usepackage{threeparttable}
\usepackage[cmex10]{amsmath}
\usepackage{amsmath}
\usepackage[english]{babel}
\usepackage{mathbbold}
\usepackage{mathtools, nccmath}
\usepackage[linesnumbered,ruled]{algorithm2e}
\makeatother
\usepackage[table]{xcolor}
\usepackage{makecell}
\usepackage{stackengine}

\usepackage{stackengine}

\usepackage{stfloats}

\ifCLASSINFOpdf
\else
\fi

\begin{document}
			\title{AI-Native Network Slicing  for  6G Networks}
\author{\small Wen Wu,~\IEEEmembership{\small Member,~IEEE,}
	Conghao Zhou,~\IEEEmembership{\small Student Member,~IEEE,}
	Mushu Li,~\IEEEmembership{\small Student~Member,~IEEE,}	 \\ 
	Huaqing Wu,~\IEEEmembership{\small Member,~IEEE,}
		Haibo  Zhou,~\IEEEmembership{\small Senior~Member,~IEEE,}
	Ning  Zhang,~\IEEEmembership{\small Senior~Member,~IEEE,}\\
Xuemin~(Sherman)~Shen,~\IEEEmembership{\small Fellow,~IEEE}, and
	Weihua~Zhuang,~\IEEEmembership{\small Fellow,~IEEE}
	\thanks{W. Wu, C. Zhou, M. Li, H. Wu, X. Shen, and W. Zhuang are with the Department of Electrical and Computer Engineering, University of Waterloo, Waterloo, ON N2L 3G1, Canada (email:\{w77wu, c89zhou, m475li, h272wu,  sshen, wzhuang\}@uwaterloo.ca). \emph{Corresponding author: Mushu Li}; }
	\thanks{H. Zhou is with the School of Electronic Science and Engineering, Nanjing University, Nanjing 210023, China (email: haibozhou@nju.edu.cn);}
	\thanks{N. Zhang is with the Department of Electrical and Computer Engineering, University of Windsor, Windsor, ON N9B 3P4, Canada (email: ning.zhang@uwindsor.ca).}}
\maketitle

\begin{abstract}
With the global roll-out of the fifth generation (5G) networks, it is necessary to look beyond 5G and envision the 6G networks. The 6G networks are expected to have space-air-ground integrated networks, advanced network virtualization, and ubiquitous intelligence. {This article presents an artificial intelligence (AI)-native network slicing architecture for 6G networks to enable the synergy of AI and network slicing, thereby facilitating intelligent network management and supporting emerging AI services. AI-based solutions are first discussed across network slicing lifecycle to intelligently manage network slices, i.e., \emph{AI for slicing}.} Then, network slicing solutions are studied to support emerging AI services by constructing AI instances and performing efficient resource management, i.e., \emph{slicing for AI}. Finally, a case study is presented, followed by a discussion of open research issues that are essential for AI-native network slicing in 6G networks.

\vspace{0.2 cm}
\begin{IEEEkeywords}
6G, AI-native, network slicing, AI for slicing, slicing for AI, ubiquitous intelligence.
\end{IEEEkeywords}
\end{abstract}

\section{Introduction}
{Compared with existing wireless networking including the fifth generation (5G), 6G is more than an improvement of key performance indicators (KPI) requirements, such as increased data rates, enhanced network capacity, and low latency. The 6G networks are envisioned to have the following unique features. First, space networks, e.g., low earth orbit (LEO) satellites, air networks, e.g., unmanned aerial vehicles (UAVs), and ground networks, e.g., cellular base stations (BSs), are integrated into a \emph{space-air-ground integrated network (SAGIN)} to provide global coverage and on-demand services~\cite{zhang2017software}. Second, resource virtualization using network slicing techniques and end user virtualization using digital twin techniques can facilitate \emph{advanced network virtualization} to provide flexible network management~\cite{shen2021Holistic, minerva2020digital}.\footnote{{End user virtualization is to virtualize end users in network operation and management by characterizing users' behaviours and status (e.g., service demands and QoS satisfaction), which can be achieved using digital twin concepts.}} Third, intelligence penetrates every corner of networks, ranging from end users, the network edge, to the remote cloud, which results in \emph{ubiquitous intelligence}. A number of network nodes are endowed with built-in artificial intelligence (AI) functionalities, thereby not only facilitating intelligent network management but also fostering AI services, e.g., deep neural network based applications. Hence, 6G networks are expected to create a new wireless networking ecosystem that brings societal and economic benefits.} 

The 6G networks will support a diverse set of services with different quality of service (QoS) requirements, such as multisensory extended reality and hologram video streaming. To support diversified services as established in 5G networks, network slicing is a potential approach to construct multiple logically-isolated virtual networks (i.e., slices) for different services on top of the common physical network~\cite{shen2020ai}. The QoS requirements of different services can be guaranteed via cost-effective slice management strategies ranging from preparation,  planning, and operation phases in the network slicing lifecycle.   

{Developing network slicing schemes faces many challenges in 6G networks due to their unique features. First, managing slices over space, air, and ground network segments in the SAGIN requires judicious coordination of heterogeneous network segments. Moreover, the 6G networks need to support a variety of new services while satisfying their different and stringent QoS requirements, which further complicates slice management. Hence, it is paramount to develop intelligent slice management solutions in 6G networks. Second, fuelled by powerful computing capability and advanced AI techniques,  ubiquitous intelligence is fostering abundant AI services with new QoS requirements, such as data quality, inference accuracy, and training latency. Hence, it is necessary to construct customized network slices to support the emerging AI services in 6G networks.}

In this article, we propose an \emph{AI-native} network slicing architecture for 6G networks to facilitate intelligent network management while supporting emerging AI services. AI-native means that, as a built-in component in the network slicing architecture, AI exists not only in the software-defined networking (SDN) controller for managing network slices, but also in network slices as services for end users. Hence, the synergy of AI and network slicing in the proposed architecture is two-fold: On one hand, AI techniques can be applied to manage network slices, namely \emph{AI for slicing}. The network slicing lifecycle including preparation,  planning, and operation phases is introduced, along with specifying AI-based solutions for each phase. In addition, the detailed procedure of information exchange among end users, access points, and the SDN controller is presented; On the other hand, network slicing can be applied to construct customized network slices for various AI services, namely \emph{slicing for AI}.  Potential approaches such as AI instance construction and efficient resource management for AI services are introduced.

The remainder of this article is organized as follows. In Section~\ref{sec:AI-Native}, expected features of 6G networks are discussed, and then the {AI-native} network slicing architecture is proposed. The basic ideas of AI for slicing and slicing for AI are presented in Section~\ref{sec:AI_for_slicing} and Section~\ref{sec:slicing_for_AI}, respectively. A case study is presented in Section~\ref{sec:case study}. In Section~\ref{sec:open_issues}, the research directions are identified, followed by the conclusion in Section~\ref{sec: conclusion}.

\section{AI-Native Network Slicing for 6G Networks}\label{sec:AI-Native}

\subsection{Network Slicing}
Network slicing is an emerging technology to support diversified applications in a cost-effective manner~\cite{shen2020ai, zhuang2019sdn}. The concept of network slicing can be traced back to the late 1980s~\cite{you2021towards}. {Nowadays, network slicing is a key technology in 5G networks}, supported by network function virtualization (NFV) and SDN techniques. Specifically, NFV enables virtualized resources and network functions for flexible resource management, while SDN facilitates centralized network management for network optimization. In 5G networks, network slicing has been defined in the 3rd generation partnership project (3GPP) Release~15~\cite{kaloxylos2018survey}. Moreover, in coming 6G networks, network slicing will continue evolving and play an increasingly important role.

The basic idea of network slicing is to create multiple logically-isolated network slices on top of the common physical infrastructure, which can achieve flexible and adaptive network management. Its benefits are three-fold: 1) Multi-tenancy - Multiple virtual networks can share the common physical infrastructure, thus reducing capital expenditures in the network deployment; 2) Service isolation - Multiple slices are constructed for different services via judicious resource management, such that service level agreements of different slices can be effectively guaranteed; 3) Flexibility - Network slicing can support flexible network management, as slices can be created, modified, or deleted on-demand. 

\subsection{Features of 6G Networks}

{From 5G to 6G, it is in general expected KPI requirements to be increased by at least an order of magnitude. According to a recent white paper~\cite{rajatheva2020white}, the KPI requirements of the 6G networks include 1\;Tbps peak data rate, 20-100\;Gbps user experienced data rate, 0.1\;{ms} end-to-end latency, 10 million\;devices/km$^2$, and near 100\% coverage. Such KPI requirements demand several candidate technologies, such as THz communications and AI~\cite{you2021towards}. The 3GPP working group will discuss 6G candidate techniques by the end of 2026, and the first 6G standard is expected to debut by 2030.}

\begin{figure*}[t]
	\centering
	\renewcommand{\figurename}{Fig.}
	\includegraphics[width=0.7\textwidth]{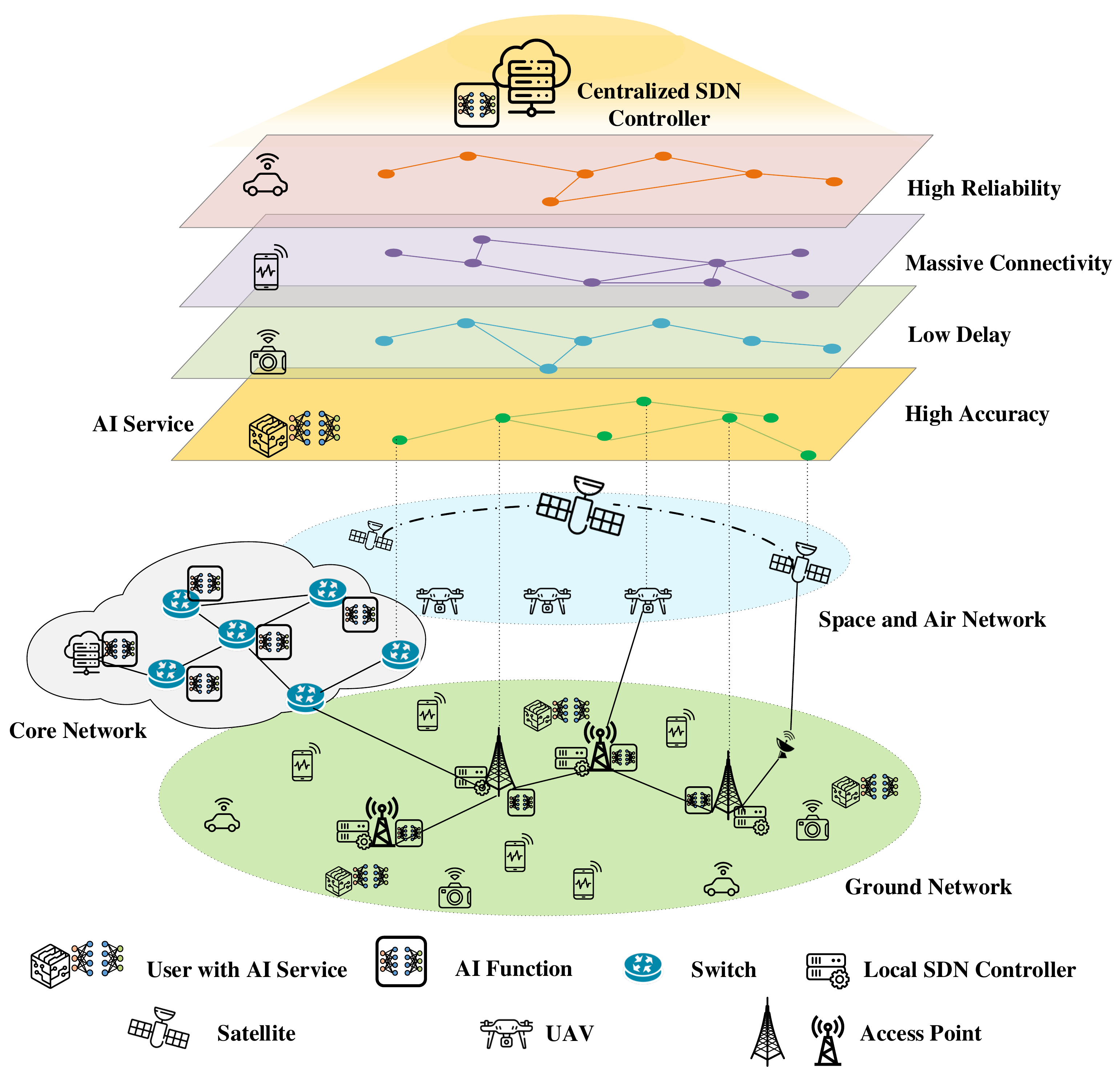}
	\caption{An illustration of the AI-native network slicing architecture for 6G networks. }
	\label{Fig:Architecture}
\end{figure*}

{Distinguished from 5G networks, 6G networks have several features:}
\begin{itemize}
		
	\item SAGIN - While current ground networks provide good coverage in highly populated areas, 6G needs to provide universal coverage, including in rural areas, remote lands, and sparsely populated areas. To achieve this goal, 6G will exploit the altitude dimension. Space, air, and ground network segments are integrated into the SAGIN~\cite{zhang2017software, zhou2020deep}, which can provide global coverage,  facilitate on-demand services, and support high-rate low-delay services;
	
	\item Diversified services  - Many services have stringent QoS requirements in different dimensions.  Mobile virtual reality (VR) and hologram video streaming applications require a high data rate, e.g., the uplink data rate of mobile VR is up to 5\;Gbps.  Other applications may require ultra-high reliability, such as autonomous driving, industrial control systems, and robot/UAV swarm, e.g., the required reliability of autonomous driving is up to 99.999\%~\cite{campolo20175g};
	
		
	\item Ubiquitous intelligence - With caching capability, a large amount of data can be stored in the network. In addition, with the development of AI techniques, edge computing, and device computing, intelligence is pushed from the remote cloud to the network edge and end users. As such, AI will be integrated into 6G networks for intelligent network management by directly learning from extensive data in the network. Moreover, ubiquitous intelligence will foster a number of AI services in which AI is provided as services.
		 
\end{itemize}



\subsection{AI-Native Network Slicing Architecture}
These features impose new challenges on developing network slicing schemes for 6G networks. {Firstly, the SAGIN not only increases the number of integrated network segments but also introduces extra dynamics on network resource availability due to satellite mobility and UAV manoeuvrability. Network slicing schemes should accommodate for the large-scale SAGIN while taking dynamic resource availability into account. Moreover, supporting diversified  services with stringent QoS requirements further complicates network slicing scheme design. Secondly, ubiquitous intelligence facilitates many emerging AI services which will be prevalent in  6G networks. Different from conventional services, facilitating AI services requires multiple steps, including collecting high-quality data samples, training satisfactory AI models, and performing low-latency model inference, which should meet diverse QoS requirements. How to satisfy such diverse QoS requirements for AI services remains a challenging issue.}

To address the above challenges, an AI-native network slicing architecture for 6G networks is presented. As shown in Fig.~\ref{Fig:Architecture}, the architecture aims at integrating SAGIN and ubiquitous intelligence and supporting diverse services with stringent QoS requirements. Compared with network slicing for 5G networks, the proposed architecture has two new characteristics. Firstly, AI is integrated into SDN controllers to realize intelligent network slicing, such that a number of network slices with stringent QoS requirements can be managed efficiently and cost-effectively via AI techniques, which is referred to as \emph{AI for slicing}. Secondly, emerging AI services are supported by network slicing. In addition to network slices for conventional services, new network slices are constructed for AI services on top of the common physical infrastructure, which is referred to as \emph{slicing for AI}. 

{Two types of SDN controllers are deployed in the proposed architecture. One is the centralized SDN controller located at the cloud, which is to manage network slices. The other is the local SDN controller located at  access points, which is to schedule resources to end users within each network slice. The SDN controller in the following refers to the centralized SDN controller unless otherwise stated.} In the following, we will illustrate the basic ideas of AI for slicing and slicing for AI in Section~\ref{sec:AI_for_slicing} and Section~\ref{sec:slicing_for_AI}, respectively.


\section{AI for Slicing}\label{sec:AI_for_slicing}
In this section, we introduce the network slicing lifecycle with three phases and then investigate potential AI solutions for each phase. Next, the corresponding procedure of  information exchange in AI for slicing is discussed.  

\subsection{Network Slicing Lifecycle}
\begin{figure*}[t]
	\centering
	\renewcommand{\figurename}{Fig.}
	\includegraphics[width=0.7\textwidth]{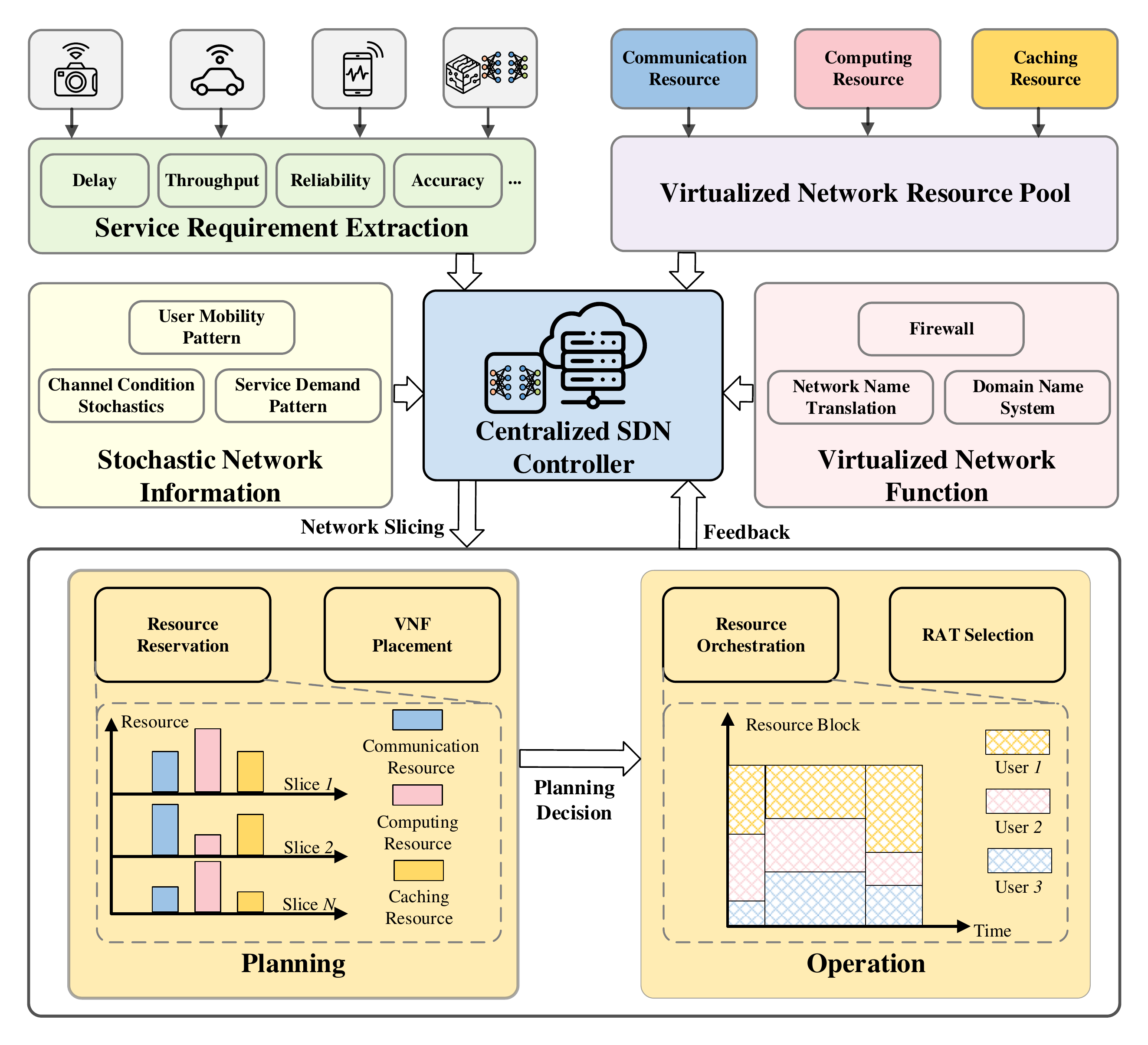}
	\caption{An illustrative example for the network slicing lifecycle {which includes preparation, planning, and operation phases.}}
	\label{Fig:AI_assited_network_slicing}
\end{figure*}

The network slicing lifecycle consists of three phases: \emph{preparation}, \emph{planning}, and \emph{operation}, {as shown in Fig.~\ref{Fig:AI_assited_network_slicing}.} {The centralized SDN controller is in charge of the preparation and planning phases, while the operation phase is coordinated by local SDN controllers.} 

\subsubsection{Preparation Phase} 
This phase is to construct and configure network slices based on service requirements, data traffic, user information, and virtual network resource availability. To achieve the goal, the SDN controller conducts the following tasks:
\begin{itemize}
	\item   Service requirement extraction - This task is to classify services by extracting their QoS requirements, such as service delay, service priority, throughput, and reliability. The 3GPP has standardized specific service/slice type values for classified services, such as enhanced mobile broadband, ultra-reliable low-latency communications, and massive machine-type communications services~\cite{kaloxylos2018survey};  
	
	\item  Network resource and function virtualization - Network resources, such as communication, computing, and caching resources, are pooled into virtualized resource blocks via advanced resource virtualization techniques. Similarly, network functions, such as firewall, network name translation, and domain name system, are separated from dedicated hardware network functions into virtualized network functions (VNFs). ﻿Through virtualization, the SDN controller can flexibly manage  network resources and  functions. 

\end{itemize} 
Once these tasks are completed, the SDN controller can construct network slices for each admitted slice request. 

\subsubsection{Planning Phase}
This phase aims at reserving network resources to slices for service provisioning. The planning phase operates in a large timescale. Time is partitioned into multiple planning periods (windows) for each slice. {The duration of each planning window depends on service demand and network dynamics, whose value ranges from several minutes to several hours.} To achieve the goal, the following two steps are conducted in the planing phase: 

\begin{itemize}
	\item Service and network  information collection - Benefiting from the global control functionality of the  SDN controller, extensive network information can be collected from underlying physical networks, such as service demands, stochastic channel conditions, and user mobility patterns. The collected information is utilized for the following resource reservation decision making;
	\item Resource reservation -  At the beginning of each planning window, the SDN controller adjusts the amount of reserved network resources for each slice based on the monitored slice performance. The reserved virtualized network resources of each slice are mapped to the physical network. At the end of each planning window, some system information is fed back to the SDN controller, such as resource utilization, system performance, and service level agreement satisfaction. Based on the feedback information, the SDN controller can adjust resource reservation decisions to accommodate dynamic network environments while guaranteeing QoS requirements.
\end{itemize}

\subsubsection{Operation Phase}
This phase is to schedule the service of a slice using the reserved resources for subscribed end users. The operation phase works in a much smaller timescale (e.g., 100\;ms) than that in the planning phase. Specifically, under the coordination of the centralized SDN controller, {local SDN controllers} allocate network resources to end users in each slice according to their real-time data traffic. The  operation decisions include selecting radio access technology (RAT), determining user association with specific radio access points, deciding proper protocol and associated parameters, and orchestrating resources among end users.



\subsection{Roles of AI in Network Slicing}

\begin{figure}[t]
	\centering
	\renewcommand{\figurename}{Fig.}
	\includegraphics[width=0.5\textwidth]{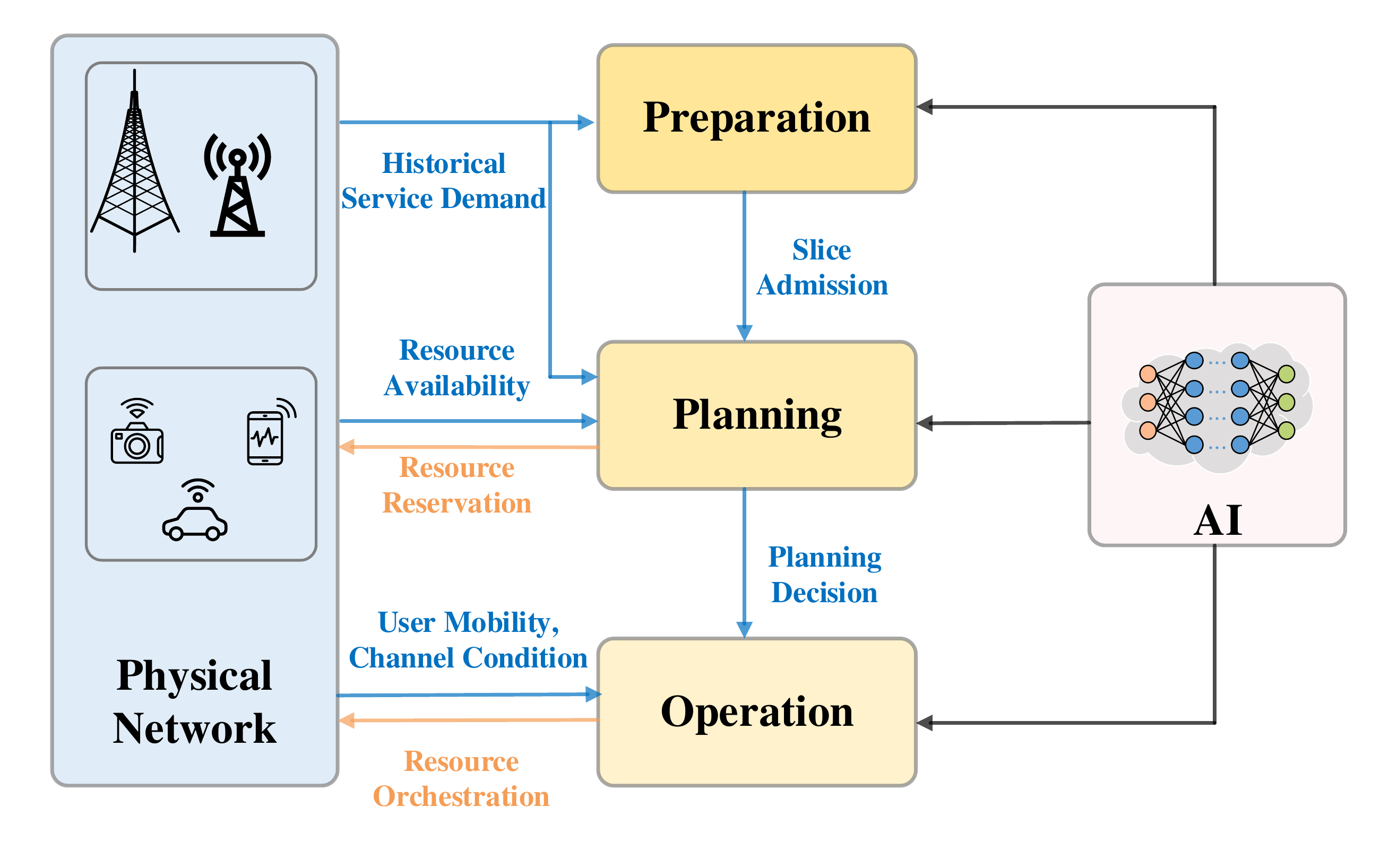}
	\caption{{The considered AI-based network slicing solution in which AI plays different roles in  preparation, planning, and  operation phases}. }
	\label{Fig:Detailed_AI_procedure}
\end{figure}

Although network slicing can facilitate service provisioning, managing a number of network slices  incurs significant network management cost, especially in 6G networks. {As shown in Fig.~\ref{Fig:Detailed_AI_procedure}, AI-based network slicing is a potential solution in which  AI  plays different roles in different network slicing phases. }

\emph{AI for  preparation}: In the preparation phase, AI needs to perform two tasks. 1) Service demand prediction - Based on historical data, service demand can be predicted via AI techniques, such as recurrent neural networks. Prior studies show that the service demand and  resource usage of a slice can  be accurately predicted~\cite{gutterman2019ran}. The prediction results can be utilized for  decision making in the planing phase. 2) Slice admission - The SDN controller admits slices to maximize network resource utilization considering resource availability and service demands. As the slice admission decision is binary, this problem is deemed as an integer optimization problem. In large-scale networks with complex resource availability distribution, conventional optimization solutions become complicated and intractable, while AI-based solutions are potential.

\emph{AI for  planning}: In the planning phase, AI can perform two tasks. 1) VNF placement - The SDN controller deploys VNFs to support services in the network. The resources allocated for VNFs should be dynamically adjusted for time-varying service demands to guarantee service delay requirements. Deep learning methods can be applied to enhance resource utilization in dynamic network environments. 2) Resource reservation - The SDN controller reserves resources for different slices based on their service demands. Since data traffic loads are time-varying, the resource reservation should be adaptive to dynamic real-time demands, which can be addressed via reinforcement learning (RL) methods, such as deep deterministic policy gradient (DDPG). 

\emph{AI for operation}: Two exemplary operation tasks are as follows: 1) Resource orchestration - The reserved resources of a slice are allocated to end users. The decisions are determined based on real-time user mobility, service demands, etc. To efficiently utilize resources, RL methods can be applied for dynamic resource orchestration; 2) RAT selection - To maximize system utility, an optimal RAT is selected among multiple candidate RATs for each end user. Due to user mobility, user-perceived service performance of an RAT is stochastic. Such problem can be addressed by multi-armed bandit methods, e.g., contextual bandit.


%
%
\subsection{Procedure of Information Exchange}

\begin{figure*}[t]
	\centering
	\renewcommand{\figurename}{Fig.}
	\includegraphics[width=0.8\textwidth]{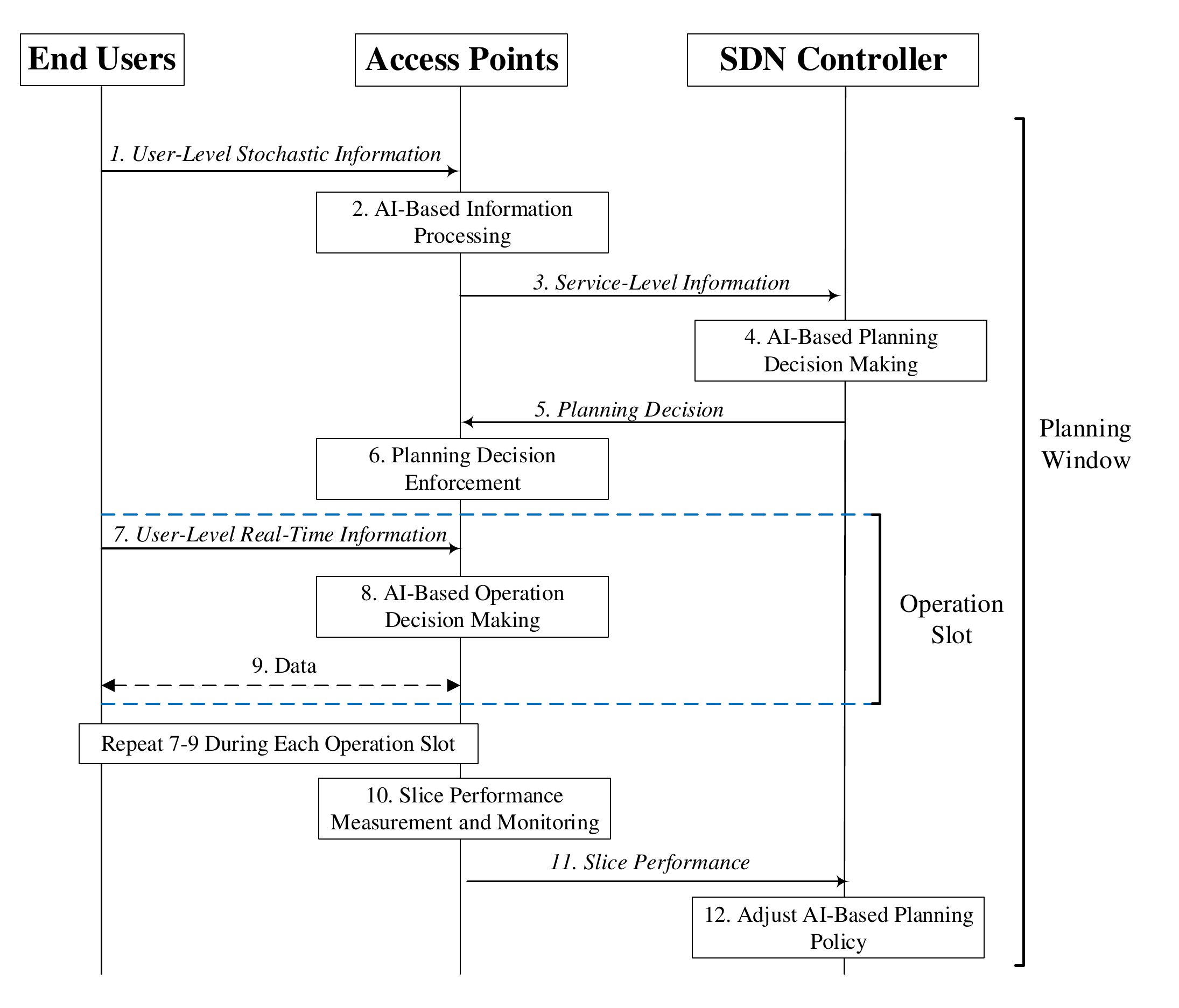}
	\caption{Procedure of information exchange in AI for slicing.}
	\label{Fig:signaling}
\end{figure*}

The AI for slicing procedure involves the information exchange among end users, access points, and the SDN controller. The procedure is illustrated in Fig.~\ref{Fig:signaling} with steps as follows:
\begin{enumerate}
	\item Access points collect user-level stochastic information, such as end users' service demand patterns, mobility patterns, and stochastic channel conditions;
	\item Access points translate the user-level information into desired service-level information. For example, user density information can be obtained from processing user location information, and AI techniques can be used for such data abstraction, fusion, and analysis; 
	\item The processed service-level information is delivered to the SDN controller;
	\item The SDN controller runs AI-based planning algorithms to make decisions based on the collected service-level information;
	\item The determined planning decisions are sent back to all access points;
	\item  Access points enforce the received planning decisions, e.g., reserving network resources for corresponding slices;
	\item End users in service report their real-time information to their associated access points, such as real-time service demands, channel conditions, and task data sizes; 
	\item  Access points run the AI-based operation algorithm to allocate resources for end users based on real-time user-level information; 
	\item Service requests from end users are supported with the allocated network resources. For example,   computation tasks can be offloaded to access points using communication resource and then processed using computing resources. For each operation slot within a planning window, Steps 7-9 are repeated;
	\item Access points monitor slice performance in the network given the enforced planning  decisions by measuring end users' satisfaction rates across all operation slots within a planing window; 
	\item Access points report network performance to the SDN controller;
	\item The SDN controller makes the planing decision for next planing window and adjusts the planning policy based on the feedback information. 
\end{enumerate} 
In the preceding procedure,  Steps 1-12 are in the network planning phase, and Steps 7-9 are in the operation phase.

\section{Slicing for AI}\label{sec:slicing_for_AI}
The slicing for AI is to utilize network slicing to support AI services while satisfying QoS requirements. Potential solutions include constructing and selecting AI instances and efficient resource management in the AI service lifecycle. 


\subsection{AI Instance}
There are diversified implementation options for supporting AI services.  An AI service can be implemented via different kinds of algorithms, training manners, and  network resource allocation. For example, objective detection services can be implemented via ResNet32, Inception-v3, AlexNet, or VGG16 algorithms. Hence, the primary issue of supporting an AI service is to determine an appropriate implementation option in the network.

\begin{figure*}[t]
	\centering
	\renewcommand{\figurename}{Fig.}
	\includegraphics[width=0.7\textwidth]{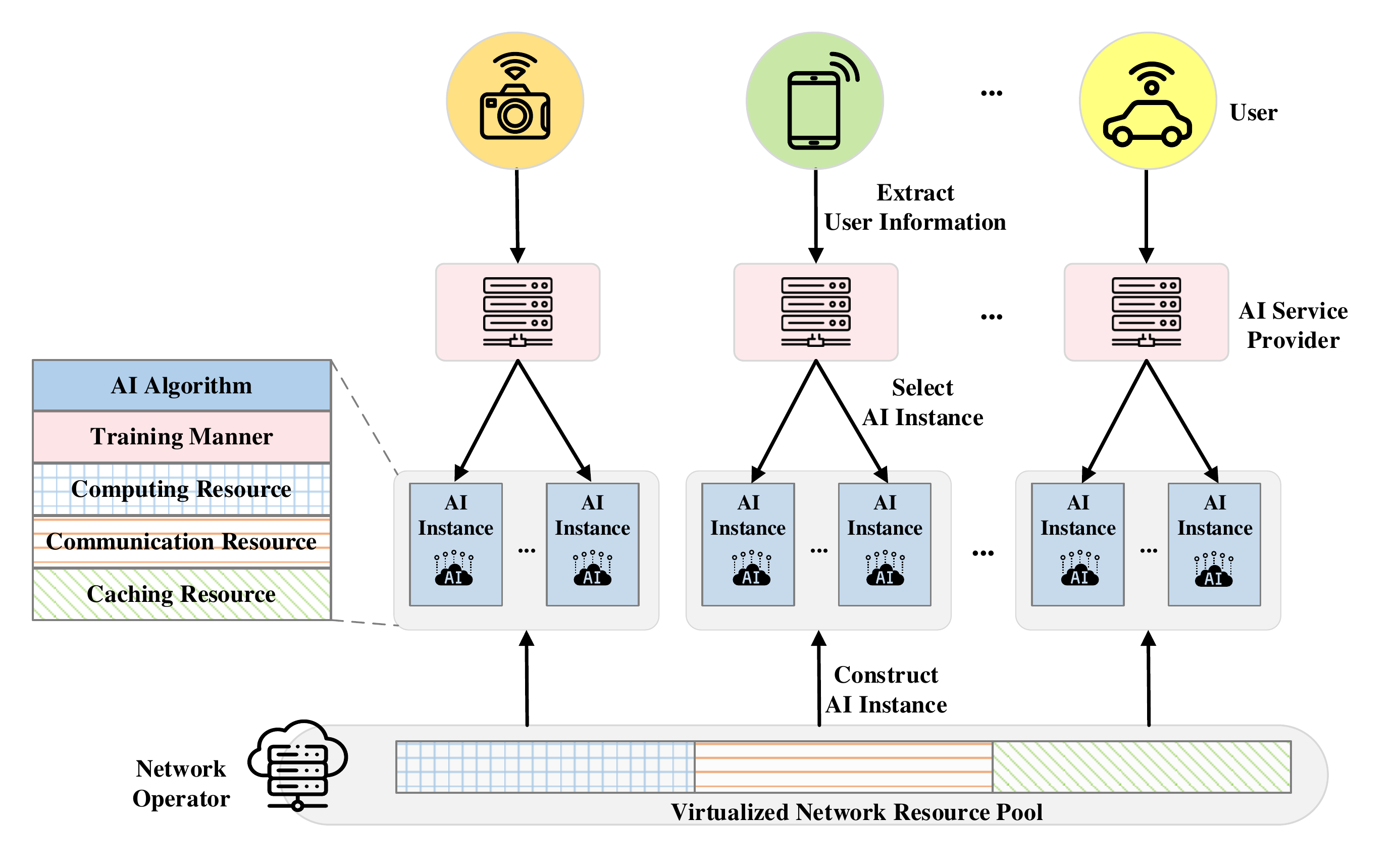}
	\caption{The conceptual AI instance management framework for AI services.}
	\label{fig:AI_instance}
\end{figure*}

We introduce the concept of \emph{AI instance} to address the issue, as shown in Fig.~\ref{fig:AI_instance}. {An AI instance of an AI service represents an implementation option for an AI service.} The basic idea is to construct multiple candidate implementation options and then select an appropriate one based on network environments. The procedure of the conceptual AI instance management framework consists of two steps. 1) AI instance construction - The network operator constructs multiple candidate AI instances for each AI service based on available virtualized network resources and AI service requirements. An AI instance may include (i) the AI algorithm which specifies the implementation algorithm and the corresponding neural network architecture, (ii) the training manner of the AI algorithm, e.g., centralized or distributed training, and (iii) the amount of the required network resources. 2) AI instance selection - In this step, the AI service provider selects an appropriate AI instance. For this purpose, the AI service provider observes the status and collects service requirements of its current subscribed end users, and then selects an AI instance among candidate AI instances provided by the network operator. {If an AI instance is selected, the AI service will be executed using the AI algorithm and the corresponding required amount of network resources by the AI instance. }In summary, the idea of AI instance provides flexibility for AI service management. 

\subsection{Resource Management in AI Service Lifecycle}
{Running AI services includes three stages: data collection, model training, and model inference, i.e., AI service lifecycle~\cite{goodfellow2016deep, mushu_VT}. Specifically, \emph{data collection} is to collect data via communication links, and the collected data can be stored in network edge servers.  Based on the collected data, an AI model can be trained in the \emph{model training} stage. The model training can be implemented in either a centralized or a distributed manner. For example, multiple devices can work collaboratively to train a global model via federated learning.  Next, well-trained AI models are deployed to execute specific computation tasks, which is referred to as \emph{model inference}. The model inference can be performed in multiple manners. For example, device-edge collaborative inference approaches can allocate and process computation tasks at different network nodes to achieve a low inference latency.}

The performance of an AI service depends on all the three stages in the AI service lifecycle. For example, model inference accuracy depends on multiple factors, such as the quality of the collected data, the number of training iterations, and the approach of model inference. Meanwhile, all these three stages consume multi-dimensional network resources. As a result, to optimize the performance of AI services, network resources should be jointly allocated for these three stages. The reserved network resources  in AI slices should be further allocated to these three stages to satisfy their corresponding QoS requirements. 

\section{Case Study}\label{sec:case study}
{In this section, a case study is provided on AI-assisted resource reservation, aiming at reducing  long-term overall system cost.}

\subsection{Considered Scenario}
{We consider an air-ground integrated network for providing autonomous driving services to vehicles traversing a highway segment. For the considered highway segment with a length of 2\;km, two BSs are uniformly deployed along the highway with a separation distance of 1\;km, and one UAV is deployed in the centre hovering at a height of 100\;meters. When an autonomous vehicle is driving on the highway, extensive computation-intensive tasks are required to be processed. For prompt task processing, all access points are equipped with edge computing servers, and vehicles can associate to the nearest access point and upload their computation tasks. We consider a delay-sensitive autonomous driving service, e.g., object detection, whose delay requirement is 100\;\emph{ms} for road safety~\cite{lin2018architectural}. The data size and computation intensity of a task are set to 0.6\;Mbit and $6\times 10^8$\;cycles, respectively. The service delay is characterized by the queuing theory since task arrivals are assumed to follow a Poisson process with rate $\lambda= 1$ packet/sec. To guarantee the service delay requirement, a network slice is constructed, in which spectrum and computing resources are reserved.}

{Resource reservation decisions are made to minimize the overall system cost considering vehicle traffic dynamics. The overall system cost is defined as~$C=\sum_{t=1}^{T}\left(\omega_r C_r^t+\omega_s C_s^t+\omega_dC_d^t\right)$, which is a weighted summation of three cost components across all $T$ planning windows. 1) Resource reservation cost $C_r^t$ accounts for the amount of the reserved spectrum and computing resources at BSs at planning window $t$. The spectrum resource is allocated in a unit of subcarrier of 5\;MHz, and the computing resource is allocated in a unit of virtual machine (VM) instance with a processing rate of $10\times 10^9$ cycles/sec; 2) Slice reconfiguration cost $C_s^t$ accounts for the difference between two consecutive resource reservation decisions~\cite{wu2020dynamic};  3) Delay requirement violation penalty $C_d^t$ refers to the penalty once service delay exceeds the delay requirement. These weight parameters are set to $\omega_r=1$, $\omega_s=20$, and $\omega_d=200$, respectively. The planning window size is set to one hour. }

{We propose a DDPG-based solution to minimize the overall system cost~\cite{wu2020dynamic}. In this solution, both actor and critic networks are fully-connected neural networks with four layers, the numbers of neurons in two hidden layers are 128 and 64, respectively, and their learning rates are set to $2\times10^{-4}$ and $2\times10^{-3}$, respectively. For performance comparison, we adopt an optimization-based solution, named \emph{myopic resource reservation}, in which network resources are reserved to minimize the resource reservation cost at each planning window while satisfying the delay requirement. }


\subsection{Simulation Results}
\begin{figure}[t]
	\centering
	\renewcommand{\figurename}{Fig.}	
	\begin{subfigure}[Convergence performance]{
			\label{Fig:One-hour system cost}
			\includegraphics[width=0.45\textwidth]{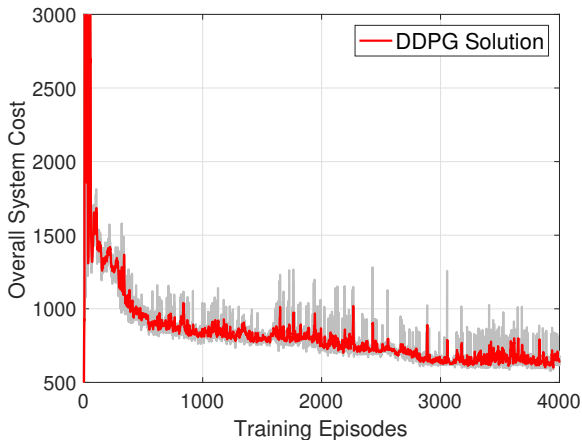}}
	\end{subfigure}
	~
	\begin{subfigure}[Cumulative system cost within one day]{
			\label{Fig:simul}
			\includegraphics[width=0.45\textwidth]{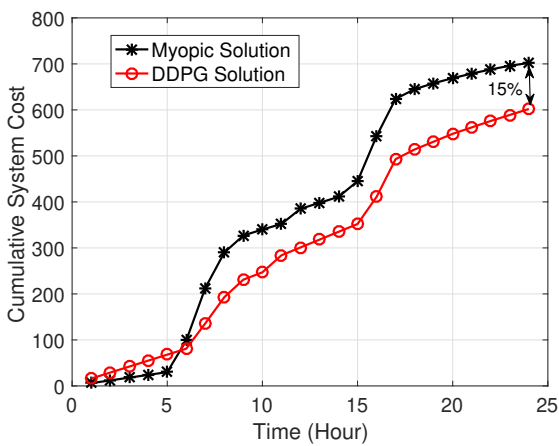}}
	\end{subfigure}
	\caption{{Performance evaluation of the proposed DDPG-based resource reservation solution.}}
	\label{Fig:Performance comparison}
\end{figure}

{We evaluate the performance of the proposed DDPG-based solution based on real-world highway vehicle traffic flow trace collected by Alberta Transportation.\footnote{Alberta Transportation: http://www.transportation.alberta.ca/mapping/.} As shown in Fig.~\ref{Fig:One-hour system cost}, we first present the convergence performance of the proposed DDPG-based solution.  A five-point moving average is applied to process raw simulation points to highlight the convergence trend (i.e., red curve).  It can be seen that the DDPG-based resource reservation solution has converged after 4,000 training episodes.}

{Next, as shown in Fig.~\ref{Fig:simul}, the cumulative system cost within one day is presented. It can be observed that the DDPG-based solution can reduce the cumulative overall system cost within one day by around 15\% as compared to the myopic solution. The reason is that the proposed DDPG-based solution is able to minimize the long-term overall system cost, while the myopic solution minimizes the short-term system cost, which incurs prohibitive slice reconfiguration cost due to frequent adjustment of network resource reservation in highly dynamic vehicular networks. The simulation results show that the proposed AI-based resource reservation solution can achieve a low system cost.}



\section{Open Research Issues}\label{sec:open_issues}
In the following, we discuss some open research issues pertaining to AI-native network slicing.

\subsection{Joint Design of Network Planning and Operation}
Planning and operation are performed and coupled in two different timescales. Specifically, the planning phase is performed at a large timescale (e.g., minute level) to reserve resources for different slices based on service demands, while the operation phase is performed at a small timescale (e.g., sub-second level) to allocate the reserved resources to on-demand users within each slice. Achieving the optimal network slicing performance requires a joint optimization design of planning and operation. 

\subsection{Data Management Framework}
The cornerstone of AI-native network slicing is abundant data that can be used for AI model training. In 6G networks, data is widely distributed in the network. Due to limited communication resources, the cost of collecting a large amount of data cannot be neglected. In addition, the collected data is required to be processed to mine valuable information for network management. For example, abundant historical behaviour data from individual users can be analyzed to predict spatio-temporal service demand distributions. Hence, establishing a data management framework to collect and analyze data is necessary.  

\subsection{Prediction-Empowered Network Slicing}
With the development of advanced AI technologies, the data traffic in the network can be predicted. How to effectively leverage the power of prediction for network slicing is an interesting topic. Since the prediction is imperfect, the prediction error may degrade the performance of network slicing. How to evaluate the impact of prediction errors on system performance and to develop corresponding solutions are  important research issues. 

\section{Conclusion}\label{sec: conclusion}
In this article, we have proposed the AI-native network slicing architecture to facilitate intelligent network management and support AI services in 6G networks. {The architecture aims at enabling the synergy of AI and network slicing. The AI for slicing is to help reduce network management complexity, while adapting to dynamic network environments by exploiting the capability of AI in network slicing. The slicing for AI is to construct customized network slices to better accommodate various emerging AI services.} To accelerate the pace of AI-native network slicing architecture development, extensive research efforts are required, such as in the identified research directions.

\section*{Acknowledgements}
This work was financially supported by Natural Sciences and Engineering Research Council (NSERC) of Canada. 
The authors would like to thank Jie~Gao, Qihao~Li, and Kaige Qu for many valuable discussions and suggestions throughout the work.


\begin{thebibliography}{10}
\providecommand{\url}[1]{#1}
\csname url@samestyle\endcsname
\providecommand{\newblock}{\relax}
\providecommand{\bibinfo}[2]{#2}
\providecommand{\BIBentrySTDinterwordspacing}{\spaceskip=0pt\relax}
\providecommand{\BIBentryALTinterwordstretchfactor}{4}
\providecommand{\BIBentryALTinterwordspacing}{\spaceskip=\fontdimen2\font plus
\BIBentryALTinterwordstretchfactor\fontdimen3\font minus
  \fontdimen4\font\relax}
\providecommand{\BIBforeignlanguage}[2]{{%
\expandafter\ifx\csname l@#1\endcsname\relax
\typeout{** WARNING: IEEEtran.bst: No hyphenation pattern has been}%
\typeout{** loaded for the language `#1'. Using the pattern for}%
\typeout{** the default language instead.}%
\else
\language=\csname l@#1\endcsname
\fi
#2}}
\providecommand{\BIBdecl}{\relax}
\BIBdecl

\bibitem{zhang2017software}
N.~Zhang, S.~Zhang, P.~Yang, O.~Alhussein, W.~Zhuang, and X.~Shen, ``Software
  defined space-air-ground integrated vehicular networks: {Challenges} and
  solutions,'' \emph{IEEE Commun. Mag.}, vol.~55, no.~7, pp. 101--109, 2017.

\bibitem{shen2021Holistic}
X.~Shen, J.~Gao, W.~Wu, M.~Li, C.~Zhou, and W.~Zhuang, ``Holistic network
  virtualization and pervasive network intelligence for {6G},'' \emph{submitted
  to IEEE Commun. Surveys Tuts.}, 2021.

\bibitem{minerva2020digital}
R.~Minerva, G.~M. Lee, and N.~Crespi, ``{Digital twin in the IoT context: A}
  survey on technical features, scenarios, and architectural models,''
  \emph{Proc. IEEE}, vol. 108, no.~10, pp. 1785--1824, Oct. 2020.

\bibitem{shen2020ai}
X.~Shen, J.~Gao, W.~Wu, K.~Lyu, M.~Li, W.~Zhuang, X.~Li, and J.~Rao,
  ``{AI}-assisted network-slicing based next-generation wireless networks,''
  \emph{IEEE Open J. Veh. Technol.}, vol.~1, no.~1, pp. 45--66, 2020.

\bibitem{zhuang2019sdn}
W.~Zhuang, Q.~Ye, F.~Lyu, N.~Cheng, and J.~Ren, ``{SDN/NFV-empowered future IoV
  with enhanced communication, computing, and caching},'' \emph{Proc. IEEE},
  vol. 108, no.~2, pp. 274--291, 2020.

\bibitem{you2021towards}
X.~You \emph{et~al.}, ``{Towards {6G} wireless communication networks: Vision,
  enabling technologies, and new paradigm shifts},'' \emph{Sci. China Inf.
  Sci.}, vol.~64, no.~1, pp. 1--74, 2021.

\bibitem{kaloxylos2018survey}
A.~Kaloxylos, ``A survey and an analysis of network slicing in {5G} networks,''
  \emph{IEEE Communications Standards Magazine}, vol.~2, no.~1, pp. 60--65,
  2018.

\bibitem{rajatheva2020white}
N.~Rajatheva \emph{et~al.}, ``{White paper on broadband connectivity in 6G},''
  \emph{arXiv:2004.14247}, 2020, [Online]. Available:
  https://arxiv.org/abs/2004.14247.

\bibitem{zhou2020deep}
C.~Zhou, W.~Wu, H.~He, P.~Yang, F.~Lyu, N.~Cheng, and X.~Shen, ``Deep
  reinforcement learning for delay-oriented {IoT} task scheduling in
  space-air-ground integrated network,'' \emph{IEEE Trans. Wireless Commun.},
  vol.~20, no.~2, pp. 911--925, 2021.

\bibitem{campolo20175g}
C.~Campolo, A.~Molinaro, A.~Iera, and F.~Menichella, ``{5G} network slicing for
  vehicle-to-everything services,'' \emph{IEEE Wireless Commun.}, vol.~24,
  no.~6, pp. 38--45, 2017.

\bibitem{gutterman2019ran}
C.~Gutterman, E.~Grinshpun, S.~Sharma, and G.~Zussman, ``{RAN resource usage
  prediction for a 5G slice broker},'' in \emph{Proc. ACM MobiHoc}, Catania,
  Italy, 2019.

\bibitem{goodfellow2016deep}
I.~Goodfellow, Y.~Bengio, and A.~Courville, \emph{Deep learning}.\hskip 1em
  plus 0.5em minus 0.4em\relax MIT press, 2016.

\bibitem{mushu_VT}
M.~Li, J.~Gao, C.~Zhou, X.~Shen, and W.~Zhuang, ``Slicing-based {AI} service
  provisioning on network edge,'' \emph{IEEE Veh. Technol. Mag.}, 2021, to
  appear.

\bibitem{lin2018architectural}
S.-C. Lin, Y.~Zhang, C.-H. Hsu, M.~Skach, M.~E. Haque, L.~Tang, and J.~Mars,
  ``The architectural implications of autonomous driving: Constraints and
  acceleration,'' in \emph{Proc. ASPLOS}, 2018, pp. 751--766.

\bibitem{wu2020dynamic}
W.~Wu, N.~Chen, C.~Zhou, M.~Li, X.~Shen, W.~Zhuang, and X.~Li, ``Dynamic {RAN}
  slicing for service-oriented vehicular networks via constrained learning,''
  \emph{IEEE J. Sel. Areas Commun.}, vol.~39, no.~7, pp. 2076--2089, 2021.

\end{thebibliography}
\end{document}